\documentclass[onecolumn,useAMS,usenatbib]{mn2e}
\usepackage{amssymb}
\usepackage{times}

\input{tcilatex}
\begin{document}

\title[MOND as a result of changing inertia]{Recovering modified Newtonian
dynamics by changing inertia}
\author[Ling Jun Wang]{Lingjun Wang\thanks{%
E-mail: initapp@hotmail.com} \\
20/F, Building 128, Nanhuxiyuan, Chaoyang District, Beijing, China}
\date{}
\maketitle

\begin{abstract}
Milgrom's modified Newtonian dynamics (MOND) has done a great job on
accounting for the rotation curves of a variety of galaxies by assuming that
Newtonian dynamics breaks down for low acceleration typically found in the
galactic contexts. This breakdown of Newtonian dynamics may be a result of
modified gravity or a manifest of modified inertia. The MOND phenomena are
derived here based on three general assumptions: 1) Gravitational mass is
conserved; 2) Inverse-square law is applicable at large distance; 3)
Inertial mass depends on external gravitational fields.

These assumptions not only recover the deep-MOND behaviour, the accelerating
expansion of the universe is also a result of these assumptions. Then
Lagrangian formulae are developed and it is found that the assumed universal
acceleration constant $a_{0}$ is actually slowly varying by a factor no more
than 4. This varying `constant' is just enough to account for the
mass-discrepancy presented in bright clusters.
\end{abstract}

\pagerange{\pageref{firstpage}--\pageref{lastpage}} \pubyear{2011}

\label{firstpage}


\section{Introduction}

The modified Newtonian dynamics (MOND), originally proposed by \citet{mond}
as an alternative to the cold dark matter paradigm to account for the
rotation curves of spiral galaxies, has extended its success to dwarfs, low
surface brightness galaxies (LSB) and ellipticals 
\citep[see][for a
review]{mond-review}. When confronting with clusters, especially rich
clusters, MOND shows some drawback. On the cluster scale, MOND still needs
dark matter, which is what MOND was particularly devised to eliminate. To
overcome this difficulty, neutrinos were speculated to be responsible %
\citep{neutrino, bullet-neutrino, neutrino-G}. Neutrinos with mass $\sim 2%
\unit{eV}$, marginally allowed by current most accurate neutrino mass
measurement, contributing negligibly to galaxies' mass budget, could be
dynamically significant in clusters of galaxies. Though this hypothesis is
successful in some aspects, it is still controversial %
\citep[e.g.][]{Pointecouteau, Angus08}.

Despite this drawback, MOND has drawn much attention because of its
impressive success compared with the standard cold dark matter paradigm,
which is facing with some difficulties, especially on the galactic scale.
MOND is a phenomenological theory that may be interpreted in different ways.
First of all, it may indicate a breakdown of Newtonian gravity %
\citep{mod-grav} where the standard Poisson equation is replaced by $\mathbf{%
\nabla }\cdot \left[ \mu \left( \left\vert \nabla \varphi \right\vert
/a_{0}\right) \mathbf{\nabla }\varphi \right] =4\pi G\rho $, and $a_{0}$ ($%
\sim 1.2\times 10^{-8}\unit{cm}\unit{s}^{-2}$), introduced by MOND, is a new
acceleration constant below which dynamics and/or gravity become
significantly non-Newtonian. \citet{mond-m} reviewed this interpretation
which relates the gravity to a potential flow. Although this is a field need
more investigation, the lack of profound physical foundation makes this
interpretation less attractive. A second interpretation is that
gravitational constant increases when accelerations are lower than $a_{0}$ %
\citep{mond-B}. This relativistic extension of MOND can mimic MOND's
behaviour at low acceleration extreme, but it is still a subject of debate %
\citep[e.g.][]{GR-confirm}.

A third interpretation is what \citet{mond} proposed that the Newtonian
dynamics may break down at low accelerations. Instead of the usual $\mathbf{F%
}=m\mathbf{a}$, \citet{mond} suggested a modified dynamics

\begin{equation}
\mathbf{F}=m_\mathrm{g}\mu \left( a/a_{0}\right) \mathbf{a,}  \label{mod-dyn}
\end{equation}

\[
\mu \left( x\gg 1\right) \approx 1,\allowbreak \allowbreak \qquad \mu \left(
x\ll 1\right) \approx x, 
\]%
where $m_{\mathrm{g}}$ is the gravitational mass of the body moving in the
field. This relation is equivalent to\qquad 
\begin{equation}
a\approx \left( a_{\mathrm{N}}a_{0}\right) ^{1/2}  \label{mod-a}
\end{equation}%
in the deep-MOND regime, where $a_{\mathrm{N}}$ is the acceleration derived
from Newtonian dynamics. It is just this simple relation that works
remarkably well in reproducing the dynamics of a variety of galaxies with
quite different morphologies and luminosities.

The modified dynamics can be interpreted as a modification of inertia,
which, if applied properly, may solve a lot of puzzles faced with modern
physics. Modification of inertia is not a new idea since the time of Mach
who challenged Newton's idea about inertia. Since the theorization of Unruh
radiation \citep{Unruh} and the discovery of Casimir effect \citep{Casimir},
their relation to inertia and gravity has frequently been speculated by
several authors \citep[e.g.][]{zpf-g, zpf-I}. Though Casimir effect is a
reality both theoretically and experimentally, what does it mean for and how
to apply it to cosmology remains a subject of debate. Puthoff's gravity,
based on Casimir effect, has the flaw of missing experimental support. As
for Unruh radiation, it is not clear what this radiation means for cosmology
and whether it is related to inertia.

In this paper, we propose a new approach, based on some general
speculations, to modification of inertia. By this modification of inertia, $%
\left( \ref{mod-a}\right) $ is successfully reproduced. Lagrangian formulae
are developed and its indications are discussed then.

\section{Assumptions and Results}

\subsection{Assumptions}

In this section, three assumptions are proposed: 1) Gravitational mass is
conserved; 2) Inverse-square law is applicable at large distance; 3)
Inertial mass depends on external fields.

If we make a comparison between gravity and electromagnetic interaction, we
immediately realise that gravitational mass is analogous to electric charge.
It is the electric charge, as a source, who produces electric field.
Electric charge cannot be created and destroyed, it is just a being.
Yang-Mills gauge theory, the foundation for standard model in particle
physics, is based on the conservation of charge. As a source of
gravitational field, it is unphysical that gravitational mass is not
conserved.

In electromagnetic field, the inverse-square law is directly related to the
zero-mass-ness of photons. Coulomb's law has been tested from $\sim 2\times
10^{10}\unit{m}$ down to $10^{-18}\unit{m}$, a magnitude span of $28$ orders %
\citep{isl-test}. The astronomical tests of Newton's gravity has been mainly
confined within solar system. The most accurate astronomical tests are
lunar-laser-ranging studies of the lunar orbit, which do not show a
deviation of gravity from Newton's law. On galactic scale, a Yukawa-like
gravity has been proposed by many authors 
\citep[see, e.g.][and references
therein]{mond-review} to account for the mass discrepancy in galaxies.
However, \citet{mond} pointed out that this is inconsistent with the
empirical Tully-Fisher law \citep{TF-law}. This line of arguments indicate
that gravity is not Yukawa-like from submillimeter scale to at least
galactic scale. Large scale structure of the Universe favors a Newton's
gravity law even on cosmic scale. Therefore it is quite safe to assume that
inverse-square law is accurate on the scale of our interest in this paper.

The equivalence of gravitational mass to inertial mass, upon which general
relativity is based, has been tested experimentally with very high accuracy %
\citep{Will}. But these tests are confined within solar system, no direct
test is available on the galactic scale. Unlike the gravitational mass,
inertial mass does not associate with any physical field. Physically,
because of the association with gravitational field, gravitational mass
should be conserved, but this is not necessarily the case for inertial mass.
Inertia is one object's ability to keep its original state of motion.
However, how does one object know its original motion state if no reference
is available?

Mach speculated that inertia is the result of the object's motion relative
to the mean mass distribution of the Universe as a whole. In other words,
inertia is meaningless if no mass is there other than the object itself. Let
us consider another situation: If the mass is distributed uniformly
throughout the Universe but one distinct object. Whatever the state of
motion of this distinct object, the state of the Universe will keep the
same. This is indicative of the dependence of inertia on the mass
distribution of the Universe.

It has been long that Milgrom found that external field plays a role on the
internal dynamics of open clusters where the internal field is well below
the critical acceleration $a_{0}$ and therefore should be in deep-MOND
regime. However, the dynamics of these systems does not show any evidence of
dark matter. Milgrom realised that external field, if significantly above
the transitional acceleration between Newtonian dynamics and MOND, could
play a role and make the dynamics of these open clusters Newtonian. In
addition, according to Mach, gravitational field, being the result of the
mass distribution of the Universe, therefore endows the object with inertia.
This is slightly different from the original Mach's Principle. Other than
simultaneously depending on the mass distribution of the whole Universe,
here we speculate that inertial mass depends on the local gravitational
field. This speculation is based both on Milgrom's realisation and on the
present day basic belief that there cannot exist action-on-distance.

The above lines of argument indicate that if no external gravitational field
exists, the object could be in random states of motion, i.e. the inertial
mass is zero. With the increase of external field, inertial mass increases
accordingly. The climbing-up of inertia, however, does not continue
infinitely. Having the external field, i.e. the gradient of the potential,
as a reference, the object has a sense of its past motion state. The
stronger the external field, the more sense it has about its original motion
state. But if the external field is strong enough, increasing the field's
strength will not increase its sense of past motion state because it just
has enough \textquotedblleft information\textquotedblright\ about its
original state of motion. As a result, we set the inertial mass $m_{\mathrm{I%
}}$ in the strong field limit to be its gravitational mass $m_{\mathrm{g}}$,
as we know from the dynamics of solar system.

However, in what a way the external gravitational field tells the moving
particle of its past motion state? The only way that this can be achieved is
by the interaction of the particle's historical trajectory with the particle
itself. That is to say, some present-day unrecognised field (referred to as
S field hereafter) is released when the particle is in motion. If the S
field is freely moving without any interruption, it will never interact with
the particle who released it. In this case, the particle does not know its
past motion state and is therefore inertialess. But if the S field it
released is interrupted, the particle will acquire inertia because of its
interaction with its prior history.

This argument seems still to be somewhat speculative. Let us now discuss a
real situation where an inertialess object could naturally acquire inertia
as a result of its interaction with its prior history. It is well known that
crystal is never perfect and full of varying kinds of defects, e.g. line
defects and point defects. Line defects are frequently referred to as cracks
in fracture mechanics. The motion of a crack is determined by the stress
field in the material under consideration. Based on a linear approximation,
the classic equation of motion for a crack is beautifully described by the
so-called linear elastic fracture mechanics (LEFM) \citep{Freund98}. LEFM
predicts that cracks are inertialess if the medium is unbounded, which has
been experimentally verified by \citet{Goldman10}.

On the other hand, if the medium is finite in size, Goldman et al. show that
the motion of a crack is influenced by the boundaries of the medium so as to
acquire inertia. This is possible because when a crack moves, the elastic
waves surrounding the crack were reflected by the boundaries and have an
influence on its motion. In other words, a moving crack interacts with its
prior history. Goldman et al. further demonstrate that the inertia of a
crack increases with crack velocity and becomes effectively infinite as the
crack's velocity approaches a limiting velocity, $c_{R}$, viz. the speed of
Rayleigh wave in this medium. This increase of inertia parallels the mass
divergence of particles in special relativity. As a result, the speed of the
proposed S field should be equal to the speed of light.

Then how a particle acquire inertia with the existence of external
gravitational field? If we treat gravitational field as \textquotedblleft
boundaries\textquotedblright\ of the medium as for the case of crack, then a
particle will acquire inertia in a similar way. The S field surrounding the
moving particle will be reflected by gravitational field so as to influence
the motion of the particle. If gravitational field vanishes, the particle
will be inertialess because the particle is moving in a unbounded
\textquotedblleft medium\textquotedblright . The stronger the gravitational
field, the more the S field is reflected so that the particle's inertia is
larger. When the gravitational field is strong enough that all S field is
reflected, the particle's inertia will not increase and stays constant.

What happens if the gravitational field vanishes and particles are therefore
inertialess? Do they move at a velocity which is larger than the speed of
light? As we said before, the S field should be at a velocity that is equal
to the speed of light. When the inertialess particles' velocity approaches
the speed of light, they will be strongly interacted by the S field they
just transmitted so that they acquire inertia. In this limit the particles
all behave like photons. Therefore we conclude that the speed of light is
still the limiting speed.

It should be stressed that inertia is only indirectly dependent on
gravitational field. What gives a particle inertia is the interaction of the
particle with the S field it transmitted previously. This point should be
borne in mind when we derive gravitational field equation in Section \ref%
{SectLagrangian}.

By these three assumptions, we shall derive the MOND relation $\left( \ref%
{mod-a}\right) $ in next section.

\subsection{Results}

Now we consider one particle's motion under the gravitational interaction of
a massive object at large separation. The particle moves inward by
converting potential energy to kinetic energy: $\mathbf{F}=\mathrm{d}\mathbf{%
p}/\mathrm{d}t=\left( \mathrm{d}m_{\mathrm{I}}/\mathrm{d}t\right) \mathbf{v}%
+m_{\mathrm{I}}\left( \mathrm{d}\mathbf{v}/\mathrm{d}t\right) $, where $%
\mathbf{p}=m_{\mathrm{I}}\mathbf{v}$ and $m_{\mathrm{I}}$, of course, is its
inertial mass, as usual. Because $\mathbf{F}=-GMm_{\mathrm{g}}\mathbf{r}%
/r^{3}$, where $M$ is the gravitational mass of the massive object, we get

\begin{equation}
m_{\mathrm{I}}\mathbf{a}+\frac{\mathrm{d}m_{\mathrm{I}}}{\mathrm{d}t}\mathbf{%
v}=-\frac{GMm_{\mathrm{g}}}{r^{3}}\mathbf{r.}  \label{general-e}
\end{equation}%
For radial motion, we have

\begin{equation}
m_{\mathrm{I}}vv^{\prime }+m_{\mathrm{I}}^{\prime }v^{2}=-\frac{GMm_{\mathrm{%
g}}}{r^{2}},  \label{fund}
\end{equation}%
here the primes denote derivatives relative to the distance $r$.
Substituting $u$ for $1/r$, the above equation is reduced to

\begin{equation}
m_{\mathrm{I}}v\frac{\mathrm{d}v}{\mathrm{d}u}+\frac{\mathrm{d}m_{\mathrm{I}}%
}{\mathrm{d}u}v^{2}=GMm_{\mathrm{g}}.  \label{start-e}
\end{equation}%
To move further on, we have to figure out, under the guide of the third
assumption proposed in above section, how $m_{\mathrm{I}}$ is shaped by
external field. In general, based on the assumption that $m_{\mathrm{I}%
}\propto m_{\mathrm{g}}$, an inertial mass function of the form

\begin{equation}
m_{\mathrm{I}}=\nu \left( \frac{g_{\mathrm{N}}}{a_{\mathrm{I}}}\right) m_{%
\mathrm{g}}  \label{inertial-mass-general}
\end{equation}%
is expected, where $\nu \left( x\right) $ is a function of $g_{\mathrm{N}%
}/a_{\mathrm{I}}$ only. As is well known, $m_{\mathrm{I}}$ is a constant in
the strong field case, and by the third assumption, $m_{\mathrm{I}%
}\rightarrow 0$ in the weak-field extreme. We are interested in the
weak-field extreme case. A plausible assumption for the weak field case is $%
m_{\mathrm{I}}\propto \left( g_{\mathrm{N}}\right) ^{\alpha }m_{\mathrm{g}}$%
, or

\begin{equation}
m_{\mathrm{I}}\left( g_{\mathrm{N}}\ll a_{\mathrm{I}}\right) =\left( \frac{%
g_{\mathrm{N}}}{a_{\mathrm{I}}}\right) ^{\alpha }m_{\mathrm{g}},
\label{inertial-mass}
\end{equation}%
where $g_{\mathrm{N}}=GM/r^{2}=GMu^{2}$, the Newtonian gravitational field.
Here we introduce a new constant, $a_{\mathrm{I}}$, which is related to
Milgrom's constant $a_{0}$, as can be seen in the next section. In order to
smoothly bridge the strong field case and weak field case, we expect $%
0<\alpha <1$, which means $m_{\mathrm{I}}$ increases rapidly when $g_{%
\mathrm{N}}\ll a_{\mathrm{I}}$ but ceases to climb up when $g_{\mathrm{N}%
}\simeq a_{\mathrm{I}}$. The index $\alpha $ has to be fixed
phenomenologically. As can be easily checked, if we set $\alpha =1/2$, the
desired MOND behaviour is recovered, i.e.

\begin{equation}
m_{\mathrm{I}}\left( g_{\mathrm{N}}\ll a_{\mathrm{I}}\right) =\left( \frac{%
g_{\mathrm{N}}}{a_{\mathrm{I}}}\right) ^{1/2}m_{\mathrm{g}},
\label{m-inertia}
\end{equation}%
where

\begin{equation}
a_{\mathrm{I}}=v_{0}^{4}/GM,  \label{a0}
\end{equation}%
and $v_{0}$ the particle's asymptotic velocity. Because $\nu \left( x\right)
\approx 1$ when $x\gg 1$, a simple but quite plausible assumption is

\begin{equation}
\nu \left( x\right) =\left( \frac{x}{1+x}\right) ^{1/2}  \label{nu_x}
\end{equation}%
for all $x$. It should be stressed that, in this model, $m_{\mathrm{I}}/m_{%
\mathrm{g}}$ is influenced only by external fields, not by any other
factors, including the particle's dynamical quantities, e.g. its velocity.
That is to say, $m_{\mathrm{I}}$ is a true scalar. Because of this behaviour
of inertial mass, the particle's Lagrangian can be expressed as equation $%
\left( \ref{lagrangian}\right) $.

If the particle is on circular orbits, therefore a constant inertial mass,
the equation of motion, by $\left( \ref{general-e}\right) $, is reduced to

\[
m_{\mathrm{g}}g_\mathrm{N}=m_{\mathrm{I}}a, 
\]%
which reads

\[
a=\left( g_{\mathrm{N}}a_{\mathrm{I}}\right) ^{1/2}, 
\]%
i.e. the recovery of equation $\left( \ref{mod-a}\right) $ if $a_{\mathrm{I}%
}=a_{0}$ is recognised. This equation, however, cannot be applied to any
other orbits other than circular ones due to the variation of the particle's
inertia. In general, equation $\left( \ref{general-e}\right) $ should be
used. This indicates that the original MOND prescription is most suitable to
describe the dynamics of spiral galaxies where internal motion is almost
perfectly circular.

Now let us consider how to escape the gravity of a massive object.
Substituting equation $\left( \ref{m-inertia}\right) $ into equation $\left( %
\ref{start-e}\right) $ we find, for a radial motion,

\begin{equation}
uv\frac{\mathrm{d}v}{\mathrm{d}u}+v^{2}=v_{0}^{2}.  \label{radial-mot}
\end{equation}%
This equation tells us that if the particle is moving away from the massive
object and has a velocity $v=v_{0}$, then $\mathrm{d}v/\mathrm{d}u=0$ and
the particle keeps moving at a constant velocity. Therefore the particle is
able to escape the gravitational pull of a massive object by itself.

If $v<v_{0}$, then $\mathrm{d}v/\mathrm{d}u>0$ and the particle decelerates
until reach a maximal distance. Of particular interest is the case of $%
v>v_{0}$, where $\mathrm{d}v/\mathrm{d}u<0$ and the particle keeps
accelerating and the acceleration is increasing. It is straightforward to
check that its velocity satisfies the following equation

\[
\left( v^{2}-v_{0}^{2}\right) ^{1/2}=Ar. 
\]%
Obviously, if no other field's disturbance, the velocity will increase
infinitely so that $v\gg v_{0}$ and

\begin{equation}
v=Ar.  \label{Hubble}
\end{equation}%
This is the direct result of the decreasing inertia while the particle moves
away from the gravitational field. This equation is formally same as
Hubble's law. But please bear in mind that Hubble's law describes velocity
field, equation $\left( \ref{Hubble}\right) $, on the other hand, describes
one particular particle's velocity change with distance. we shall defer the
discussion of this equation to Section \ref{s-unbound}.

It is clear that, since the particle's mass is dependent on the strength of
gravity, gravity's apparent effect is not always attractive, but takes on
different aspects dependent on the particle's motion state. If the particle
is bound to the massive object, the gravity's apparent effect is attraction.
On the other hand, if the particle is in an unbound state, the gravity's
effect is repulsion, as indicated by $\left( \ref{Hubble}\right) $, the
particle's velocity is not decreasing but increasing as it moves away from
the gravitational field.

\subsection{Lagrangian Formalism\label{SectLagrangian}}

In above section we just assume that inertial mass depends on the external
field and then extend Newtonian dynamics in a minimum manner. To formulate a
self-consistent theory we need to develop a set of Lagrangian formulae. As
usual, we write down the Lagrangian:

\begin{equation}
L=\frac{1}{2}m_{\mathrm{I}}v^{2}-m_{\mathrm{g}}\phi ,  \label{lagrangian}
\end{equation}%
where $\phi $ is the scalar gravitational field that is determined by
Poisson equation $\mathbf{\nabla }^{2}\phi =4\pi G\rho $ with $\rho $ the
gravitational mass density\footnote{%
As stated above, when we vary $\phi $, we should get Possion equation
because $m_{\mathrm{I}}$ is not directly dependent on gravitational field.}.
The Euler-Lagrange equations of motion

\[
\frac{\mathrm{d}}{\mathrm{d}t}\left( \frac{\partial L}{\partial \dot{x_{i}}}%
\right) -\frac{\partial L}{\partial x_{i}}=0 
\]%
give

\begin{equation}
\frac{\mathrm{d}\mathbf{p}}{\mathrm{d}t}=-m_{\mathrm{g}}\mathbf{\nabla }\phi
+\frac{1}{2}v^{2}\mathbf{\nabla }m_{\mathrm{I}}.  \label{motion}
\end{equation}%
Compared with equation $\left( \ref{general-e}\right) $ we find that
equation $\left( \ref{motion}\right) $ contains an additional term $\frac{1}{%
2}v^{2}\mathbf{\nabla }m_{\mathrm{I}}$, stemming from the fact that inertial
mass depends on external field, so is a function of position. It turns out
that this term plays a key rule in accounting for the dynamics of dwarfs,
spiral galaxies and clusters in a consistent and self-contained way.
Applying $\mathbf{p}=m_{\mathrm{I}}\mathbf{v}$ to the left hand side of
equation $\left( \ref{motion}\right) $ gives:

\begin{equation}
m_{\mathrm{I}}\mathbf{a}=m_{\mathrm{g}}\mathbf{g}_{\mathrm{N}}+\frac{1}{2}%
v^{2}\mathbf{\nabla }m_{\mathrm{I}}-\left( \mathbf{\nabla }m_{\mathrm{I}%
}\cdot \mathbf{v}\right) \mathbf{v.}  \label{motion-e}
\end{equation}%
For the deep-MOND case, substituting equation $\left( \ref{m-inertia}\right) 
$ into above equation yields

\begin{equation}
\mathbf{a=}\left( \frac{a_{\mathrm{I}}}{g_{\mathrm{N}}}\right) ^{1/2}\mathbf{%
g}_{\mathrm{N}}+\frac{1}{4}\frac{v^{2}}{g_{\mathrm{N}}}\mathbf{\nabla }g_{%
\mathrm{N}}-\frac{1}{2}\frac{\mathbf{\nabla }g_{\mathrm{N}}\cdot \mathbf{v}}{%
g_{\mathrm{N}}}\mathbf{v.}  \label{deep-mond-general}
\end{equation}%
This equation is the general motion equation for the deep-MOND regime,
including the non-spherically symmetric systems. It is apparent from
equation $\left( \ref{motion-e}\right) $ that in a modified inertia theory,
unlike modified gravity, the acceleration experienced by particle depends
not only on position, but on the velocity as well. This seems a drawback of
this kind of theory, but on the other hand this could be an unparalleled
advantage \citep[see, e.g.,][]{mond-m}.

As above, let us first consider the circular motion. In this case the above
equation, after applying equation $\left( \ref{m-inertia}\right) $, reduces
to:

\begin{equation}
a=\left( a_{\mathrm{I}}g_{\mathrm{N}}\right) ^{1/2}\left( 1+\frac{1}{2}\frac{%
v^{2}}{\sqrt{GMa_{\mathrm{I}}}}\right) .  \label{circular-a}
\end{equation}%
From this equation we can easily find the stable circular velocity

\begin{equation}
v_{\mathrm{c}}=\left( 4GMa_{\mathrm{I}}\right) ^{1/4}.  \label{max-v}
\end{equation}%
However, only the sufficiently virialised systems can attain to this high
circular velocity. For those systems less virialised, the objects on the
outskirts of the systems will follow quasi-circular orbits and gradually
wind up during the inward migration. This indicates that Milgrom's constant
is actually slowly varying according to

\begin{equation}
a_{0}=a_{\mathrm{I}}\left( 1+\frac{v^{2}}{v_{\mathrm{c}}^{2}}\right) ^{2},
\label{varying-a0}
\end{equation}%
that is to say, $a_{0}$ is varying in a narrow range $a_{\mathrm{I}}\leq
a_{0}\leq 4a_{\mathrm{I}}$ and older systems tend to have higher values of $%
a_{0}$.

To find the value of $a_{\mathrm{I}}$, an easy but reliable way is to
carefully select a large enough sample of well studied galaxies so that the
scatter of $a_{0}$ is moderate and MOND works quite well for this sample. we
select the galaxies from \citet{BBS}, \citet{Sanders96} and %
\citet{Sanders98b}. The reasons for selecting these galaxies are three-fold.
First, these galaxies are among the highest quality of observational data
and it was demonstrated that MOND can account for the data with a relatively
high precision. Secondly, these galaxies are all spiral galaxies for which
MOND is most successful. In addition, as indicated above, $a_{0}$ evolves
with galaxies. To reduce the scatter of $a_{0}$, we should select galaxies
with comparable properties. Finally, these three samples contain several
galaxies, e.g. NGC 2841, that are quite controversial within the context of
MOND. It is quite desirable if a varying $a_{0}$ can settle down these
issues. The resulting sample contains 63 spiral galaxies, but 9 galaxies
from \citet{Sanders98b} are eliminated from the list adopted to calculate $%
a_{0}$ because of the reason presented below.

In Table 1 the calculated values of $a_{0}$ are listed for every galaxy in
the sample. $a_{\mathrm{I}}$ is found by requiring the average value of $%
a_{0}=1.21\times 10^{-8}\unit{cm}\unit{s}^{-2}$:

\begin{equation}
a_{\mathrm{I}}=0.667\times 10^{-8}\unit{cm}\unit{s}^{-2}.  \label{a_I}
\end{equation}%
Several points should be mentioned. The accelerations at the last measured
points of the rotation curves $\left( a_{\mathrm{lmp}}\right) $ of NGC 3949,
NGC 3953, NGC 4085, UGC 6973 are in the Newtonian regime and therefore
eliminated from the calculation of $a_{\mathrm{I}}$. In light of the found
value of $a_{\mathrm{I}}$, we expect that the galaxies with $a_{\mathrm{lmp}%
} $ comparable to or larger than $a_{\mathrm{I}}$ will be quite dynamically
different from other galaxies, these galaxies are NGC 3877, NGC 3972, NGC
4051, NGC 4217, NGC 4389, with $a_{\mathrm{lmp}}>0.7\times 10^{-8}\unit{cm}%
\unit{s}^{-2}$. As a result, these five galaxies are eliminated from the
calculation of $a_{\mathrm{I}}$. But it should be mentioned that, if these
five galaxies are included in the galaxies responsible for calculating $a_{%
\mathrm{I}}$, $a_{\mathrm{I}}$ would be slightly smaller. By this selection
criterion, the resulting scatter of $a_{0}$, $\pm 0.30$, around the mean
value is moderate, as desired.

\subsection{Comments on Selected Individual Galaxies\label{s-comment}}

A quick glance of the result for $a_{0}$ suggests that large and high
surface brightness galaxies tend to have a high $a_{0}$. Several comments
are presented below for those values significantly deviated from the
standard value:

\textit{NGC 2841.} -- This is the most controversial spiral galaxy for MOND. %
\citet{BBS} found that, if a Hubble distance $\left( 9.5\unit{Mpc}\right) $
was used, MOND obviously fails to account for the rotation curve. On the
other hand, if the Tully-Fisher distance was used, MOND is in well agreement
with the observed data. However, the Tully-Fisher distance is twice as the
Hubble distance, a seemly unacceptable result. Subsequently, %
\citet{Cepheid-distance} found a distance of $14.1\unit{Mpc}$ to NGC 2841
based on Cepheid method. This distance goes half way between the Hubble
distance and Tully-Fisher distance. As a result, the situation is alleviated
somewhat, but the predicted curve still deviates systematically from the
observed data. Table 1 gives this galaxy's $a_{0}$ a value of $1.48\times
10^{-8}\unit{cm}\unit{s}^{-2}$, the largest in this sample if not include
the values not adopted for calculating $a_{\mathrm{I}}$, which is more than
22\% larger than the standard value. This value will further alleviate the
situation and make MOND in principle compatible with the data. But it should
be pointed out that even with this high value, the data still can not be
comfortably matched by the prediction. Here is a caveat. We find that, for
this galaxy, $a_{\mathrm{lmp}}=0.66\times 10^{-8}\unit{cm}\unit{s}^{-2}\sim
a_{\mathrm{I}}$. Because $a_{\mathrm{I}}$ is the transition acceleration
beyond which the modified inertia transits to Newtonian inertia, we expect
the value $a_{0}$ calculated in this way is only qualitatively correct. If
otherwise $a_{\mathrm{lmp}}\ll a_{\mathrm{I}}$, the value listed in Table 1
should be exact. But NGC 2841 evolved beyond this stage because it has a
much higher $a_{\mathrm{lmp}}$. Quantitatively, a value of $1.87\times
10^{-8}\unit{cm}\unit{s}^{-2}$ for $a_{0}$ will do the work if the Cepheid
distance is used. If we adopt equation $\left( \ref{nu_x}\right) $ as the
inertial mass dependence on external gravitational field, we find a value of 
$1.86\times 10^{-8}\unit{cm}\unit{s}^{-2}\left( \sim 2.77a_{\mathrm{I}%
}\right) $ for $a_{0}$, which is just what is needed for this galaxy to
bring MOND prediction in accordance with the observation. As in the general
case, when $a_{\mathrm{lmp}}\sim a_{\mathrm{I}}$, the actual $a_{0}$ depends
on the ratio of rotation velocity to its stable circular velocity. When
rotation velocity approaches its stable circular velocity, $a_{0}$ will have
a value of $2.45\times 10^{-8}\unit{cm}\unit{s}^{-2}\left( \sim 3.66a_{%
\mathrm{I}}\right) $. This implies $2.77a_{\mathrm{I}}\leq a_{0}\leq 3.66a_{%
\mathrm{I}}$ when $a_{\mathrm{lmp}}\sim a_{\mathrm{I}}$. We find that the
maximum value $\left( 3.66a_{\mathrm{I}}\right) $ for $a_{0}$ when $a_{%
\mathrm{lmp}}$ is in the vicinity of $a_{\mathrm{I}}$ is less than the
corresponding value $\left( 4a_{\mathrm{I}}\right) $ when $a_{\mathrm{lmp}%
}\ll a_{\mathrm{I}}$, indicating that the Newtonian dynamics begin to take
over when acceleration enters the Newtonian regime. The oscillating rotation
curve, rather than a perfect flat rotation curve, of this galaxy on the
outskirts may have some implication of this subtlety.

\textit{DDO 154.} -- A value of $0.9\times 10^{-8}\unit{cm}\unit{s}^{-2}$,
significantly less than the average value, is given for this dwarf and
gas-rich galaxy. Nonetheless, this low value can be largely attributed to
the low rotation velocity at the last measured points. This is a quite
controversial case for MOND. \citet{DDO154-M} demonstrated that this galaxy
is among the most acute test of MOND. Shortly, \citet{Lake89}, however,
showed that a substantially small value of $a_{0}$ is necessary to account
for the observed curve. \citet{Milgrom-91} disputed Lake's conclusion,
arguing that a different distance to this galaxy be more reliable. But if
the ideas presented in this paper are correct, a small value of $a_{0}$ is
inevitable because of the declining nature of the rotation curve at the
outermost points. The internal acceleration of DDO 154 is everywhere less
than $a_{0}$ so that MOND should play a dominant role. But MOND cannot
account for the declining nature of the rotation curve. In the spirit of
this paper, however, the declining nature of gas-rich dwarf galaxies can be
naturally accounted for. The main point is that gas-rich galaxy is newly
formed object with a less relaxed outer part and a fairly relaxed inner
part. Consequently, the inner part should have a higher value of $a_{0}$
than its outer part. This indicates that $a_{0}$ is varying not only from
galaxy to galaxy, but also within galaxies. For sufficiently relaxed
galaxies, $a_{0}$ could be treated as a constant without loss of accuracy,
but for gas-rich galaxies, the variation of $a_{0}$ is maximised.

\textit{NGC 3893.} -- Table 1 gives $a_{0}$ a value of $1.47\times 10^{-8}%
\unit{cm}\unit{s}^{-2}$. This high value is largely the result of the high $%
a_{\mathrm{lmp}}$. But because this galaxy has a disturbed velocity field,
the result should not be taken seriously.

\textit{NGC 3992.} -- Table 1 gives $a_{0}$ a value of $1.48\times 10^{-8}%
\unit{cm}\unit{s}^{-2}$ or $1.86\times 10^{-8}\unit{cm}\unit{s}^{-2}$ if
equation $\left( \ref{nu_x}\right) $ is used, an apparently not necessary
adjustment for the standard value if the good agreement between predicted
and observed rotation curve is noticed. However, as mentioned by %
\citet{Sanders98b}, the fitted value of mass-to-light ratio of the stellar
disk is unusually large compared with other values in that sample. If,
however, an $a_{0}$ with above value is used, the near-infrared
mass-to-light ratio will be reduced by a factor of $1.22$ or $1.54$, a value
compatible to other galaxies in the sample.

\textit{NGC 4010.} -- Table 1 gives $a_{0}$ a value of $1.44\times 10^{-8}%
\unit{cm}\unit{s}^{-2}$ and even higher if equation $\left( \ref{nu_x}%
\right) $ is used. This high value of $a_{0}$ seems unexpected for such a
low surface brightness galaxy. However, this galaxy has a very extended disk
and the scale length is much larger compared with other galaxies in its
mother cluster \citep{Tully97}. As a result, this galaxy has a much larger
internal acceleration, so is $a_{0}$. This galaxy is an example that $a_{0}$
does not actually directly depend on surface brightness. Since young
galaxies with outstanding star formation activities tend to have a high
surface brightness even if its internal acceleration is low. On the other
hand, some old galaxies devoid of star formation tend to have a low surface
brightness because of their old star population. In addition, we find that
the rotation curve of this galaxy begins to decline at large radii, another
example that $a_{0}$ is varying within one galaxy.

Recently \citet{Swaters10} study a sample of 27 dwarf and low surface
brightness galaxies within the framework of MOND and find that there are
some systematics that cannot be reconciled easily: MOND curve predicts
higher rotation velocities in the outer regions and lower in the central
regions for low surface brightness dwarfs; higher rotation velocities in the
central regions and lower in the outer regions for high surface brightness
galaxies (UGC 5721, UGC 7323, UGC 7399, UGC 7603, UGC 8490). This intriguing
phenomenon forces Swaters et al. to try a fit with $a_{0}$ as a free
parameter. By allowing $a_{0}$ vary, Swaters et al. find a trend that lower
surface brightness galaxies tend to have lower $a_{0}$. This is just what
equation $\left( \ref{varying-a0}\right) $ expected if lower surface
brightness galaxies tend to be less relaxed. If equation $\left( \ref%
{varying-a0}\right) $ is applied to this sample of systems, an average value
of $1.06\times 10^{-8}\unit{cm}\unit{s}^{-2}$ is found, which is
significantly lower than the commonly used value. \citet{Swaters10},
however, find an even lower average value of $0.7\times 10^{-8}\unit{cm}%
\unit{s}^{-2}$ for $a_{0}$ in this sample, a value that is just above $a_{%
\mathrm{I}}$. Though the calculated value of $a_{0}$ is significantly lower
than the commonly used value, it is still significantly higher than the
fitted average value in this sample. Furthermore, the fitted values for the
galaxies in this sample have an enormously large scatter that is completely
out of the reach for $\left( \ref{varying-a0}\right) $ to account for. This
enormous scatter, on the one hand, is a consequence of the internal dynamics
of the systems; on the other hand, is a consequence of observational
uncertainties. If such a large scatter of $a_{0}$ is necessary for this
sample, then it will be a challenge for the ideas presented in this paper.

\subsection{Dynamical Mass of Clusters of Galaxies}

It has been quite controversial with the applicability of MOND to clusters. %
\citet{The88} first noted that MOND is inconsistent with the dynamics of
Coma cluster unless the MOND acceleration parameter $a_{0}$ is 4 times
larger than the value inferred from spiral galaxy rotation curves. %
\citet{Sanders94}, based on a small sample of X-ray-emitting clusters,
concluded that MOND mass is consistent with the detected gas mass.
Subsequently, based on a much larger sample of X-ray-emitting clusters,
Sanders found that MOND over-predicted the mass by a factor of 2 than the
detected gas mass \citep{Sanders99}. In light of this finding, the mass
over-prediction was already implied by \citet{Sanders94} though with a less
statistical significance.

This mass over-prediction flaws MOND since MOND was particularly speculated
to eliminate non-baryonic dark matter. To remedy MOND, \citet{neutrino}
suggested that neutrinos, with mass of $2\unit{eV}$, is responsible for the
mass discrepancy in rich clusters. This mass of neutrinos, having a
negligible effect on the galaxy scale, could have a significant dynamical
effect on cluster scale. Unfortunately, \citet{Angus08} showed that
neutrinos are incompatible with the observed mass distribution within X-ray
bright groups and clusters. Consequently, Angus et al. proposed that sterile
neutrinos would close the cluster mass problem. This conclusion contradicted
the finding by \citet{mond-group-mil} who found that MOND is compatible with
galaxy groups. Unfortunately, by studying the gravitational lensing of
clusters, \citet{NZ2008} conclude that even sterile neutrinos are far from
enough to close the mass budget of clusters.

However, if the inertia is allowed to vary, as presented above, no dark
matter is needed. Of the 93 clusters studied by \citet{Sanders99}, the
internal accelerations are of $\sim 0.6\times 10^{-8}\unit{cm}\unit{s}^{-2}$%
, that is, in the vicinity of $a_{\mathrm{I}}$. Section \ref{s-comment}
shows that for systems with internal accelerations comparable to $a_{\mathrm{%
I}}$, a much higher effective $a_{0}$ should be used. It can be checked that
for these 93 clusters, $a_{0}$ approaches the maximum value $3.66a_{\mathrm{I%
}}\left( 2.02\times 1.21\times 10^{-8}\unit{cm}\unit{s}^{-2}\right) $. This
high value is just enough to eliminate the mass discrepancy in bright
clusters. In this respect, we say that the dynamics of bright clusters are
quite similar to that of NGC 2841. This also implies that not all clusters
would have the mass discrepancy problem within the framework of MOND. Those
clusters with much lower internal accelerations will have similar dynamics
as spiral galaxies.

But the mass discrepancy problem is not completely closed because in some
cases MOND over-predicts masses by a factor of 3 or even higher. If this
phenomenon is ubiquitous and no conventional interpretations, e.g. dark
baryons or neutrinos, are reliable, then there really exists a dark matter
problem for MOND.

\subsection{Unbound Systems\label{s-unbound}}

After a discussion of circular motion, let us now turn to radial motion. In
this case equation $\left( \ref{motion-e}\right) $ reduces to

\[
a=\left( a_{\mathrm{I}}g_{\mathrm{N}}\right) ^{1/2}\left( 1-\frac{1}{2}\frac{%
v^{2}}{\sqrt{GMa_{\mathrm{I}}}}\right) . 
\]%
We find the conclusion is not changed much for radial motion by an
application of Lagrangian formula. The only difference is a change of the
asymptotic velocity, defined by equation $\left( \ref{max-v}\right) $ other
than the value $v_{0}$ determined by equation $\left( \ref{a0}\right) $.

Equation $\left( \ref{Hubble}\right) $ reminds us of the Hubble's law. But
does it really imply Hubble's law. By its low magnitude ($H_{0}=72\unit{km}%
\unit{s}^{-1}\unit{Mpc}^{-1}$) compared to the typical galaxies' asymptotic
rotation velocities ($\simeq 200\unit{km}\unit{s}^{-1}$) and their typical
sizes ($\lesssim 100\unit{kpc}$), if it is really Hubble's law, we expect
that Hubble's law is not the result of mass inhomogeneity on the galactic
scale. On contrary, Hubble flow is a result of mass inhomogeneity on cosmic
scale. Obviously, if the mass distribution is completely homogeneous,
galaxies would not feel the existence of an external gravitational field,
resulting in zero inertial masses and therefore random motions. But actually
every galaxy is embedded in an ubiquitous gravitational field. A fluctuation
of mass density on cosmic scale gives rise to the small gravity on large
scale, those cosmic objects unbound by this inhomogeneity of mass
distribution acceleratingly expand. If $\delta M$ is the mass inhomogeneity
caused by a density perturbation $\Delta =\delta \rho /\rho $ on the cosmic
scale, we expect the Hubble recession velocity exceeds $\delta v_{0}=\left(
4G\delta Ma_{\mathrm{I}}\right) ^{1/4}$ so that an accelerating expansion is
guaranteed. This indicates that the accelerating expansion of the Universe
can be most readily observed on scales that the clustering of galaxies
becomes negligible. The two-point correlation function of galaxies shows
that these scales are $r\simeq 5h^{-1}\unit{Mpc}$. So the accelerating
expansion of the Universe should be free of peculiar contamination beyond
these scales. The accelerating expansion of the Universe seems to imply an
increase of the universe energy, or in modern parlance, the existence of
dark energy. But above analysis shows that the accelerating expansion of
Universe does not violate conservation of energy because equation $\left( %
\ref{Hubble}\right) $ is derived based on the conservation of energy. All
difference is the dependence of inertia on gravity. Although galaxies'
velocities are becoming larger and larger, their inertial masses are
dropping dramatically, resulting in the conservation of total energy.

Although we recover MOND's relation $\left( \ref{mod-a}\right) $ in the case
of circular orbits, the ideas presented above are quite different from that
of MOND. In MOND, gravity's effect is always attractive, the moving
particles in the gravitational field are all feeling the same acceleration
as long as they locate at the same place in the field. In fact MOND,
including the relativistic MOND theory proposed by Bekenstein, complies with
weak equivalence principle. This is however not the case for the ideas
presented here. Just as we already find, gravity is attractive in the case
of bound state, and repulsive in an unbound configuration. Even for
particles in the same type of state, say bound state, their accelerations
could be quite different because of their different radial velocities, which
determine at what rate the moving particles' inertial masses change. As a
consequence, all particles moving from large distance towards the massive
object tend to converge to the same velocity, i.e. the asymptotic velocity $%
v_{0}=\left( 4GMa_{\mathrm{I}}\right) ^{1/4}$. If the particles' velocities
are less than $v_{0}$, the attraction effect of the gravity increases their
velocities. On the other hand, if the particles' velocities are larger than $%
v_{0}$, the repulsion effect of the gravity decrease their velocities. Only
in the Newtonian regime is the gravity always attractive.

In addition, since the gravitational potential is not infinite, opposed to
the original MOND, the moving body can escape the gravitational pull easily
without resorting to other nearby massive galaxies' assist as described by %
\citet{escape-mond}. This is yet another advantage of the changing inertia
proposal over the traditional MOND. In many cases, by finely tuning the
parameters, the predictions of CDM model are barely discernible from that of
MOND. But globular clusters are outliers, which are believed contain no dark
matter. Consequently \citet{BGK2005} suggested a test of gravitational
theories by studying velocity dispersion of globular clusters. Recently
these studies are carried out by several groups, finding an inconsistency
between MOND and observations \citep{MT2008,Haghi2009,SN2010}. The problem
is that MOND systematically overpredicts the velocity dispersions. This can
be understood as follows: Because of the infinite escape velocity intrinsic
in MOND, at any given radius the velocity distribution should be large. This
is no longer the case for the changing inertia proposal since at any radius
the escape velocity is finite.

\section{Discussion}

\subsection{Indications for structure formation and evolution}

The cosmology and structure formation in the context of MOND were first
studied by \citet{Felten84} and then extended by \citet{Sanders98a}, %
\citet{Sanders01} and \citet{Stachniewicz01}. The main findings of these
studies are that MOND would boost the structure formation even with a very
low matter density. \citet{van den Bosch00} semianalytically studied the
formation of disk galaxies and found that dark matter and MOND models are
indistinguishable. \citet{Stachniewicz05} found that the dark ages would end
at $z\sim 30$, in good agreement with the WMAP results. More recent studies %
\citep{Sanders08,Malekjani09} of structure formation generally strengthen
these early conclusions. When confronting MOND with the formation of galaxy
clusters, however, \citet{Nusser06} found that the simulated MOND clusters
are significantly denser than the observed ones. It means that MOND, as a
gravitational field, is so strong that it over-evolves the structure on
cluster scale.

If the ideas presented in this paper are correct, we find that the changing
inertia postulate boosts the structure formation but the evolution is only
moderately boosted. This is because when the particles begin to fall into
the potential well of the density perturbation, thanks to their low initial
velocities, therefore small variation of their inertial masses in unit time
duo to small variation of the potential well in which they are moving, the
accelerations of the particles are almost the same as that predicted by the
standard MOND. This indicates that the timescale of initial formation is
barely different from that of standard MOND. But when the particles gain
enough velocities, owing to the rapid variation of their inertial masses,
their accelerations are quite small compared with the ones predicted by the
standard MOND. Therefore the structure dynamical evolution is deferred. This
effect will make the structure less dense than that predicted by MOND. %
\citet{van den Bosch00} also found that stellar feedback is needed to
reproduce the lack of high surface brightness dwarf galaxies within the
context of MOND. However, the changing inertia proposal will defer the
evolution of structure so that central surface brightness could be
significantly lower.

This line of arguments also indicates that older objects in a system should
be kinematically hotter than their younger counterparts. It is well known
that there exists an age-velocity relation (AVR) in the disk of spiral
galaxies \citep[cf.][]{Meusinger91}. The ideas presented in this paper
indicate that there should be a rapid rise of the velocity dispersion of
disk stars and then saturate when they are old enough. Unfortunately, the
kinematical evolution of the disk is still controversial. %
\citet{Edvardsson93} and \citet{Soubiran08} found that there is a rapid rise
of velocity dispersion for the first 2 $\unit{Gyr}$ and then saturate for
older stars. \citet{Nordstrom04} and \citet{Rocha-Pinto04}, on the other
hand, found a steady increase in the velocity dispersion with age. %
\citet{Holmberg07}, with an improvement on $T_{\mathrm{eff}}$ and [Fe/H]
calibrations for early F stars, found a steady increase in velocity
dispersion with age and a tentative saturate at age 10 $\unit{Gyr}$.

The age-velocity relation is usually attributed to disk heating. Several
heating mechanisms, e.g. scattering by giant molecular clouds, spiral
structures, and accretion, have been proposed to account for this relation.
But it is growingly evident that these mechanisms are insufficient to
account for the observations \citep[cf.][]{Rocha-Pinto04}. The ideas
presented in this paper may shed new light on this issue because they will
boost the velocity dispersion.

\subsection{More words on inertia}

When we argue for a dynamical origin of inertia, we mentioned cracks and the
acquisition of their inertia as an interaction with their prior histories by
introducing boundaries for the medium. Here we present evidence that why
cracks are of particular interest. Cracks, the line defects in material,
have their siblings in crystal, viz. dislocations and disclinations. Since
the development of defect gauge field theory in condensed matter physics, it
has been more and more clear that Einstein's gravitational theory is a
special case of the gauge theory for defected crystal in the sense that
general relativity is free of torsion \citep[cf.][]{Kleinert90}. The
geometrical study of crystal finds that a crystal filled with dislocations
and disclinations has the same geometric properties as an affine space with
torsion and curvature, respectively.

The Einstein curvature tenser, $G_{\mu \nu }$, in its three-dimensional
version, is equivalent to the disclination density tenser, $\Theta _{ij}$.
As a result, the Einstein field equation, $G_{\mu \nu }=-8\pi GT_{\mu \nu }$%
, indicates that the baryons are just disclinations in the \textquotedblleft
world crystal\textquotedblright\ with Plank length as the lattice constant %
\citep{Kleinert90}. Then why do we not observe a space with torsion? Torsion
can only be a result because of the existence of dislocations. However,
dislocation would make the space discontinuous and cannot exist
macroscopically because it is unable to maintain stress equilibrium by
itself. This is, however, not to say that dislocations do not exist
microscopically. Actually the dislocation density, $\alpha _{ij}$, is
related to disclination density by $\partial _{i}\alpha _{ij}=-\varepsilon
_{jkl}\Theta _{kl}$. That is to say, several dislocations could end up as a
disclination as seen by a distant observer. Therefore dislocations are akin
to quarks in the way that dislocations and quarks are always confined within
disclinations and baryons, respectively. If quarks are indeed dislocation
defects in world crystal, then we should expect that space is not torsion
free on the quark scale. Furthermore, the conservation laws for the angular
momentum and energy-momentum density in general relativity are just the
respective conservation laws for dislocation and disclination density in
defected world crystal \citep{Kleinert90}.

This indicates that the argument for the way particles acquire inertia in a
similar way in which a crack acquires inertia in a bounded medium is not
just a trick but a facet of reality.

\section{Conclusions}

MOND has been well established as a serious alternative to the standard dark
matter model for about three decades owing to its remarkable success in
accounting for the dynamics of a variety of galaxies with quite different
luminosities, morphologies. However, when confronting with dwarfs and
clusters, MOND is controversial. There exists cumulated evidence that the
acceleration parameter $a_{0}$, assumed to be universal, is varying in the
sense that low surface brightness galaxies tend to have low $a_{0}$. For
bright clusters, a factor of 2 mass over-prediction is well established.
Again, this over-prediction can be accommodated by an $a_{0}$ 2 times larger
than the usually adopted value.

With above observational evidence, a motivation to theoretically account for
the phenomena is prompt. This paper postulates that the MOND phenomenology
can be accounted for by three assumptions: 1) Gravitational mass is
conserved; 2) Inverse-square law is applicable at large distance; 3)
Inertial mass depends on external fields. The first two assumptions are
quite general and should not find objection. The third assumption, given by
equation $\left( \ref{m-inertia}\right) $ appropriate at the low
acceleration regime, is key to reproduce the phenomenology of MOND. It is
found that the inertia modified in this way exactly recover the formulae
suitable to MOND in the circular motion case.

By a consideration of Lagrangian formulae, however, it is found that the
usual relation $\left( \ref{mod-a}\right) $ is replaced by $\left( \ref%
{circular-a}\right) $. A comparison of equations $\left( \ref{mod-a}\right) $
and $\left( \ref{circular-a}\right) $ shows that $a_{0}$, a proposed
universal constant, is actually varying in a narrow range, $a_{\mathrm{I}%
}\leq a_{0}\leq 4a_{\mathrm{I}}$, if the internal accelerations of the
systems in question (or the external field) are much lower than $a_{\mathrm{I%
}}$. A scrutiny of equations $\left( \ref{motion-e}\right) $ and $\left( \ref%
{nu_x}\right) $ indicates that the effective $a_{0}$ not only depends on the
orbital velocity but also on the external field. This varying $a_{0}$ is
just enough to eliminate the over-prediction factor of 2 found in bright
clusters and the lower value in low surface brightness galaxies. Because of
this nature of $a_{0}$, $a_{0}$ not only varies from system to system, but
also varies within one galaxy. This phenomenon can readily account for the
decline of rotation curves found for many low surface brightness galaxies.

Above statement suits only for circular motion. For radial motion, situation
is quite different. It turns out that there exists a critical velocity, $v_{%
\mathrm{c}}$ defined by equation $\left( \ref{max-v}\right) $, for every
galaxy. If the object in radial motion is moving slower than this critical
velocity, gravity appears to be attractive, as usual. But for a motion
faster than this critical velocity, gravity appears to be repulsive in the
sense that the object begins to accelerate. This immediately reminds us of
the accelerating expansion of the universe. By a careful inspection of the
cosmic parameters, if the modified inertia is correct, it is realised that
the accelerating expansion of the universe must be the result of
inhomogeneity of the universe on cosmic scales.

In general, besides circular motion and radial motion, equations $\left( \ref%
{motion-e}\right) $ and $\left( \ref{nu_x}\right) $ should be used to
account for the dynamics of systems in question. These two equations
indicate that the traditional MOND prescription is only appropriate to
describe circular motions. It is found that $a_{0}$ is not a fundamental
constant, its value is an indication of the system's age and relaxation
index. In general, high surface brightness objects tend to have high value
of $a_{0}$. But this is not appropriate for all systems. $a_{0}$ is actually
determined by the orbital velocities and external gravitational fields. A
more accurate value of $a_{\mathrm{I}}$ should be determined by these two
equations. But owing to the success of MOND in spiral systems, the true
value of $a_{\mathrm{I}}$ should not deviate far from the value given by
equation $\left( \ref{a_I}\right) $.

\section*{Acknowledgments}

I am grateful to Robert H. Sanders for his encouragement in this field. I
also wishes to thank Moti Milgrom for his help during the year the ideas
discussed here were conceived. I thank HongSheng Zhao for his critical
reading of the manuscript and helpful comments.

\bigskip

Table 1. The sample selected for the calculation of $a_{\mathrm{I}}$.
Calculated values of $a_{0}$ are listed on column 2. Reference: (1) %
\citet{BBS}; (2) \citet{Sanders96}; (3) \citet{Sanders98b}. Nine galaxies
from \citet{Sanders98b}, NGC 3877, NGC 3949, NGC 3953, NGC 3972, NGC 4051,
NGC 4085, NGC 4217, NGC 4389, UGC 6973, are eliminated from the calculation
of $a_{\mathrm{I}}$ because of the reason presented in text. The data for
UGC 6446 are taken from \citet{Swaters10} to utilise the most recent update,
but the differences are small.

\begin{tabular}{ccc}
\hline
Galaxy & $a_{0}\left( 10^{-8}\unit{cm}\unit{s}^{-2}\right) $ & ref \\ \hline
\multicolumn{1}{l}{NGC 2403} & 1.20 & 1 \\ 
\multicolumn{1}{l}{NGC 2841} & 1.48 & 1 \\ 
\multicolumn{1}{l}{NGC 2903} & 1.32 & 1 \\ 
\multicolumn{1}{l}{NGC 3109} & 1.01 & 1 \\ 
\multicolumn{1}{l}{NGC 3198} & 1.11 & 1 \\ 
\multicolumn{1}{l}{NGC 6503} & 1.06 & 1 \\ 
\multicolumn{1}{l}{NGC 7331} & 1.29 & 1 \\ 
\multicolumn{1}{l}{NGC 1560} & 1.12 & 1 \\ 
\multicolumn{1}{l}{UGC 2259} & 1.24 & 1 \\ 
\multicolumn{1}{l}{DDO 154} & 0.90 & 1 \\ 
\multicolumn{1}{l}{DDO 170} & 0.99 & 1 \\ 
\multicolumn{1}{l}{UGC 2885} & 1.28 & 2 \\ 
\multicolumn{1}{l}{UGC 5533} & 1.17 & 2 \\ 
\multicolumn{1}{l}{NGC 6674} & 1.16 & 2 \\ 
\multicolumn{1}{l}{NGC 5907} & 1.34 & 2 \\ 
\multicolumn{1}{l}{NGC 2998} & 1.20 & 2 \\ 
\multicolumn{1}{l}{NGC 801} & 1.12 & 2 \\ 
\multicolumn{1}{l}{NGC 5731} & 1.24 & 2 \\ 
\multicolumn{1}{l}{NGC 5033} & 1.24 & 2 \\ 
\multicolumn{1}{l}{NGC 3521} & 1.25 & 2 \\ 
\multicolumn{1}{l}{NGC 2683} & 1.31 & 2 \\ 
\multicolumn{1}{l}{NGC 6946} & 1.16 & 2 \\ 
\multicolumn{1}{l}{UGC 128} & 1.00 & 2 \\ 
\multicolumn{1}{l}{NGC 1003} & 0.98 & 2 \\ 
\multicolumn{1}{l}{NGC 247} & 1.22 & 2 \\ 
\multicolumn{1}{l}{M33} & 1.32 & 2 \\ 
\multicolumn{1}{l}{NGC 7793} & 1.36 & 2 \\ 
\multicolumn{1}{l}{NGC 300} & 1.09 & 2 \\ 
\multicolumn{1}{l}{NGC 5585} & 1.10 & 2 \\ 
\multicolumn{1}{l}{NGC 2915} & 1.06 & 2 \\ 
\multicolumn{1}{l}{NGC 55} & 1.15 & 2 \\ 
\multicolumn{1}{l}{IC 2574} & 1.05 & 2 \\ 
\multicolumn{1}{l}{DDO 168} & 1.14 & 2 \\ 
\multicolumn{1}{l}{NGC 3726} & 1.20 & 3 \\ 
\multicolumn{1}{l}{NGC 3769} & 1.02 & 3 \\ 
\multicolumn{1}{l}{NGC 3893} & 1.47 & 3 \\ 
\multicolumn{1}{l}{NGC 3917} & 1.33 & 3 \\ 
\multicolumn{1}{l}{NGC 3992} & 1.48 & 3 \\ 
\multicolumn{1}{l}{NGC 4010} & 1.44 & 3 \\ 
\multicolumn{1}{l}{NGC 4013} & 1.26 & 3 \\ 
\multicolumn{1}{l}{NGC 4088} & 1.38 & 3 \\ 
\multicolumn{1}{l}{NGC 4100} & 1.31 & 3 \\ 
\multicolumn{1}{l}{NGC 4138} & 1.31 & 3 \\ 
\multicolumn{1}{l}{NGC 4157} & 1.30 & 3 \\ 
\multicolumn{1}{l}{NGC 4183} & 1.11 & 3 \\ 
\multicolumn{1}{l}{UGC 6399} & 1.24 & 3 \\ 
\multicolumn{1}{l}{UGC 6446} & 1.05 & 3 \\ 
\multicolumn{1}{l}{UGC 6667} & 1.23 & 3 \\ 
\multicolumn{1}{l}{UGC 6818} & 1.17 & 3 \\ 
\multicolumn{1}{l}{UGC 6917} & 1.31 & 3 \\ 
\multicolumn{1}{l}{UGC 6923} & 1.44 & 3 \\ 
\multicolumn{1}{l}{UGC 6930} & 1.23 & 3 \\ 
\multicolumn{1}{l}{UGC 6983} & 1.15 & 3 \\ 
\multicolumn{1}{l}{UGC 7089} & 1.14 & 3 \\ \hline
\end{tabular}

\label{lastpage}

\end{document}